\documentclass[prd,eqsecnum,preprint,showpacs,nofootinbib]{revtex4}
\usepackage{bm}
\usepackage{stmaryrd}
\usepackage{amsmath}
\usepackage{amssymb}
\usepackage{graphicx}
\usepackage{textcomp}
\usepackage{calrsfs}
\usepackage{yfonts}

\newcommand{\order}{\mathcal{O}}

\newcommand{\nn}{\nonumber}

\newcommand{\be}{\begin{equation}}
\newcommand{\ee}{\end{equation}}
\newcommand{\ben}{\begin{equation*}}
\newcommand{\een}{\end{equation*}}
\newcommand{\bea}{\begin{eqnarray}}
\newcommand{\eea}{\end{eqnarray}}

\begin{document}

\title{Hard and soft walls}

\author{Kimball A. Milton}\email{milton@nhn.ou.edu}
\affiliation{Homer L. Dodge Department of
Physics and Astronomy, University of Oklahoma, Norman, OK 73019-2061, USA}
\date{\today}
\begin{abstract}
In a continuing effort to understand divergences which occur when quantum 
fields are confined by bounding surfaces, we investigate local energy 
densities (and the local energy-momentum tensor) in the vicinity of a wall.  
In this paper, attention is largely confined to a scalar field.  If the wall 
is an infinite Dirichlet plane, well known volume and surface divergences are 
found, which are regulated by a temporal point-splitting parameter.  If the 
wall is represented by a linear potential in one coordinate $z$, the 
divergences are softened.  The case of a general wall,
described by a potential of the form $z^\alpha$ for $z>0$ is considered.  
If $\alpha>2$, there are no surface divergences, which in any case vanish
if the conformal stress tensor is employed.  Divergences within the wall
are also considered.
\end{abstract}

\maketitle

\section{Introduction}
Quantum vacuum energy, or Casimir energy, referring to the quantum energies 
of fields in the presence of material bodies or boundaries, is a mature 
subject \cite{Bordag:2009zz}. In the last decade there have been tremendous 
advances in both experiment and theory, so that the so-called Lifshitz theory 
\cite{Lifshitz:1956zz} has been confirmed at the 1\% level,
and theoretically there is now the possibility of calculating forces 
between bodies of practically any shape and constitution.
Even the effects of finite temperature have now been confirmed 
\cite{Sushkov:2010cv}.  Yet, such effects are controversial, and there 
are many issues that are still unresolved.

One of the issues that has been controversial almost from the beginnings of 
the subject is that of the Casimir self-energy of an object, as opposed to 
the energy of interaction between two or more rigid objects.  For example, 
in 1968 Boyer calculated the self-energy of a perfectly conducting spherical 
shell of zero thickness and found a surprising repulsive result 
\cite{Boyer:1968uf}.  This result has been confirmed by different techniques 
by many authors since.  Yet within a decade, the meaning of this result
was profoundly questioned \cite{Deutsch:1978sc}; not only is the meaning of 
self-energy rather obscure, but divergences occur whose omission has resulted 
in controversy up to the present time \cite{miltonself}.  Some of these 
divergences are proportional to the volume, to the surface area, and to the 
corners, so-called Weyl terms, which can be unambiguously removed.  
Curvature divergences are rather more subtle, and the reason Boyer obtained 
a finite result was that the interior and exterior curvature contributions 
cancel.  Situations without curvature, such as triangular prisms 
\cite{Abalo:2010ah} and tetrahedra \cite{Abalo2011}, have finite
calculable self-energies when only the interior contributions are included.

The above calculations refer to the total energies of the systems.  Yet, 
there is much interest in local quantum  energy densities, or more generally, 
the vacuum expectation value of the stress-energy tensor.  There are 
well-known divergences in these as surfaces are approached \cite{miltonself}. 
However, most of the work on this subject has studied perfect boundaries, such 
as ideal conductors or Dirichlet walls.  Since there are still issues 
regarding surface divergences that are not well understood, which are 
particularly relevant when the coupling to gravity is considered, in this 
paper we will consider walls that are modeled by potentials that are 
``softer'' than such a perfect wall.  (Soft walls were considered earlier, but 
apparently only for the global energy \cite{actor}.) In particular,
we will consider massless scalar fields in three dimensions in the presence 
of a potential which depends only on one coordinate, $z$.  We follow Bouas 
et al.~\cite{fulling} and consider semi-infinite potentials, so that the 
potential vanishes for $z<0$ while it is a monomial in $z$ for $z>0$.  For 
special cases (Dirichlet, linear, and quadratic potentials) the energy density 
may be found explicitly in terms of known functions, but in general asymptotics
yield the information about the nature of the divergences as the region of the 
potential is approached from the left, as well as the divergences in the
region of the potential.  Unlike many previous investigations of this subject,
including some by Milton \cite{miltonbook,miltonself}, 
we precisely regulate all expressions by inserting a 
temporal point-splitting.  Then precise forms of the divergences in terms of 
the temporal splitting parameter are obtained, which exhibit the expected Weyl
terms, as well as the nature of the singularity at the boundary $z=0$.  Unlike 
Ref.~\cite{fulling} we consider a general stress tensor with arbitrary 
conformal parameter $\xi$; that reference considers $\xi=1/4$, but we find that
for $\xi=1/6$ the divergences that occur as the boundary is approached are 
removed; in any case, they disappear for a potential higher than quadratic.

\section{Dirichlet Wall}
Consider a massless scalar field in three-dimensional space subject to 
a Dirichlet wall
\be v(z)=\left\{\begin{array}{cc}
                0,&z<0,\\
                \infty,&z>0.
               \end{array}\right.
\ee
The Green's function, the solution to
\be
\left(\frac{\partial^2}{\partial t^2}-\nabla^2+v\right)G(x,x')=\delta(x-x'),
\ee
has the form
\be
G(x,x')=\int\frac{d\omega}{2\pi}\frac{(d\mathbf{k_\perp})}{(2\pi)^2 }
e^{-i\omega(t-t')}
e^{i\mathbf{k_\perp\cdot(r-r')_\perp}}g(z,z';\kappa),\label{ftgf}
\ee
where, for $z,z'<0$,
\be
g(z,z',\kappa)=-\frac1\kappa e^{\kappa z_<}\sinh \kappa z_>,\label{gf}
\ee
where
$z_<$, $z_>$ is the lesser, greater of $z$ and $z'$.  Here 
\be
\kappa^2=k_\perp^2-\omega^2,
\ee
where we have anticipated making the Euclidean rotation (not just a Wick
rotation)
\be
\omega\to i\zeta,\quad (t-t')\to i\tau,
\ee
so we may regard $\kappa$ as positive.

The energy-momentum tensor for the scalar field is
\be
t^{\mu\nu}=\partial^\mu\phi\partial^\nu\phi-\frac12 g^{\mu\nu}
\left(\partial_\lambda
\phi\partial^\lambda\phi\right)-\xi\left(\partial^\mu\partial^\nu-g^{\mu\nu}
\partial^2\right)\phi^2,
\ee
where $\xi$ the conformal parameter, which for the conformal value of 
$\xi=\frac16$
yields a traceless stress tensor, $t^\lambda{}_\lambda=0$.
The connection between the classical causal (Feynman) Green's function and the
(time-ordered) vacuum expectation values of the fields is
\be
\langle \phi(x)\phi(x')\rangle=\frac1iG(x,x'),\label{vev}
\ee
so the one-loop vacuum energy density of the field is
\be
u(z)=\langle t^{00}\rangle= \frac1{2i}\left(\partial^0\partial^{\prime0}+
\bm{\nabla\cdot\nabla'}\right)G(x,x')\bigg|_{x'\to x}+i\xi\nabla^2 G(x,x).
\ee

Using the Fourier representation (\ref{ftgf}) we have for the energy density
\be
u(z)=\frac12\int\frac{d\zeta}{2\pi}\frac{(d\mathbf{k_\perp})}{(2\pi)^2} 
e^{i\zeta\tau}
\left[\left(-\zeta^2+k_\perp^2+\frac\partial{\partial z}
\frac\partial{\partial z'}
\right)g(z,z')\bigg|_{z'\to z}-2\xi \frac{\partial^2}{\partial z^2}g(z,z)
\right].
\label{u}
\ee
Here, we have regulated the integral by retaining $\tau$ as a temporal 
point-splitting regulator, to be set equal to zero at the end of the 
calculation. We now introduce polar coordinates in the $\zeta$-$\mathbf{k}$ 
volume, so that
\be
\zeta=\kappa\cos\theta,\quad |\mathbf{k_\perp}|=\kappa\sin\theta
\ee
and then the integral over the regulator term is
\be
\int_{-1}^1d\cos\theta\, e^{i\tau\kappa\cos\theta}
=\frac2{\kappa\tau}\sin\kappa\tau.
\ee
Inserting the Green's function (\ref{gf}) into the energy integral (\ref{u}) 
yields
\be
\frac1{8\pi^2}\int_0^\infty d\kappa\,\kappa^3 \left[e^{2\kappa z}\left(4\xi-1-
\frac1{\kappa^2}
\frac{\partial^2}{\partial\tau^2}\right)+\frac1{\kappa^2}
\frac{\partial^2}{\partial\tau^2}
\right]\frac2{\kappa\tau}\sin\kappa\tau.\label{u1}
\ee

The two terms in Eq.~(\ref{u1}) consist of a $z$-dependent term and a constant.
The latter is just the bulk energy density arising from the free part of the 
Green's function,
\be
g_0(z,z')=\frac1{2\kappa}e^{-\kappa|z-z'|}.
\ee
This is evaluated as
\be
u_0=\frac1{4\pi^2}\frac{\partial^2}{\partial\tau^2}\frac1\tau\int_0^\infty 
d\kappa\sin\kappa \tau
=\frac3{2\pi^2\tau^4},
\ee
which uses the integral
\be
\int_0^\infty dx\,\sin x=1.
\ee
This result is just the well-known volume Weyl term.
If $|z|\gg\tau$, we can take $\tau\to0$ in the remaining term,
and we find for the Dirichlet wall
\be
u(z)-u_0=-\frac{1-6\xi}{6\pi^2}\int_0^\infty d\kappa\,\kappa^3 e^{2\kappa z}=
-\frac{1-6\xi}{16\pi^2 z^4}.\label{dsdiv}
\ee
This is exactly the form found near one of the plates for the two-plate
Casimir situation \cite{miltonbook,miltonself}.
Note that this term vanishes for $\xi=1/6$, which suggests
that it has no significance, disregarding gravity.  
If we keep the regulator, we can integrate
over the whole volume to the left of the wall:
\be
\int_{-\infty}^0 dz(u-u_0)=\frac1{8\pi^2}\int_0^\infty d\kappa\,\kappa^2
\left[4\xi\frac{\sin\kappa\tau}{\kappa\tau}+2\frac{\cos\kappa\tau}
{(\kappa\tau)^2}-
2\frac{\sin\kappa\tau}{(\kappa\tau)^3}\right]=-\frac1{8\pi\tau^3}.\label{W2}
\ee
This uses the evaluations
\be \int_0^\infty d\kappa \cos\kappa \tau=0,\quad \int_0^\infty d\kappa\,\kappa
\sin\kappa\tau=0,\quad \int_0^\infty \frac{d\kappa}\kappa \sin\kappa\tau 
=\frac\pi2.\label{fints}
\ee
The result (\ref{W2}) is exactly the second Weyl term, expressing the 
energy per unit area for a Dirichlet wall.

In exactly the same way we can compute all the components of the stress tensor.
The result, for $|z|\gg \tau$, is exactly as expected:
\be
\langle t^{\mu\nu}\rangle=\frac1{2\pi^2\tau^4}\mbox{diag}(3,1,1,1)+
\frac{1-6\xi}{16\pi^2z^4}\mbox{diag}(-1,1,1,0).
\ee
The bulk term has the required traceless, rotationally-invariant form,
since it is unaware of the wall. 
 The surface-divergent term vanishes for
the conformal case, and exhibits no force on the wall, so is unobservable. 
This stress tensor trivially satisfies energy-momentum conservation,
$\partial_\mu \langle t^{\mu\nu}\rangle=0$.

Incidentally, note that if the cutoff were omitted for the bulk term,
we would obtain a form that is consistent not with rotational symmetry,
but with the symmetry for the $2+1$ dimensional breakup as seen in the
finite part of the stress tensor for the interaction between two Dirichlet
plates:
\be
u_0\to -\frac1{12\pi^2}\int_0^\infty d\kappa\,\kappa^3
\mbox{diag}(1, -1, -1, 3).
\ee

\section{Linear Wall}
 
We next consider the linear wall,
\be
v(z)=\left\{\begin{array}{cc}
            0,&z<0.\\
z,&z>0.
           \end{array}\right..
\ee
The energy density to the left of the wall is given by Eq.~(\ref{u}),
whereas to the right of the wall, the potential must be included, or
in general
\begin{eqnarray}
 u(z)&=&\frac12\langle\left[
(\partial^0\phi)^2+\bm{\nabla}\phi\cdot\bm{\nabla}\phi+v\phi^2\right]
-2\xi\nabla^2\phi^2]\rangle.
\label{ed}
\end{eqnarray}
In Eq.~(\ref{ed}) the fields are to be evaluated at coincident points, and
again the connection with the Green's function is given by Eqs.~(\ref{vev})
and (\ref{ftgf}),
where now the reduced Green's function satisfies
\begin{equation}
 \left(-\frac{\partial^2}{\partial z^2}+k^2+v(z)-\omega^2\right)g(z,z')
=\delta(z-z').\label{gfe}
\end{equation}
As we saw before, it is convenient to perform a Euclidean rotation, 
$\omega\to i\zeta$.

\begin{figure}
\centering
\includegraphics{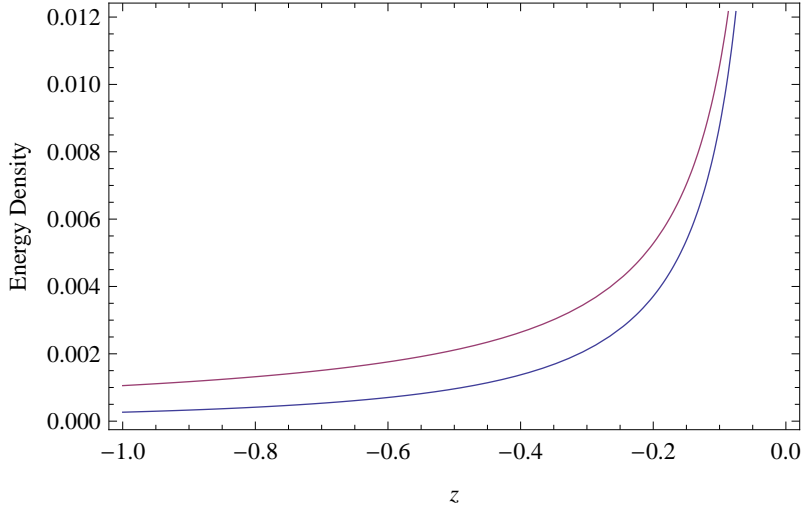}
\caption{\label{fig1}
Energy density (divided by $6\xi-1$) to the left of a linear potential.  
The exact result (lower curve) is
compared with the asymptotic behavior for small $z$, Eq.~(\ref{asym}).}
\end{figure}

\begin{figure}
\centering
\includegraphics{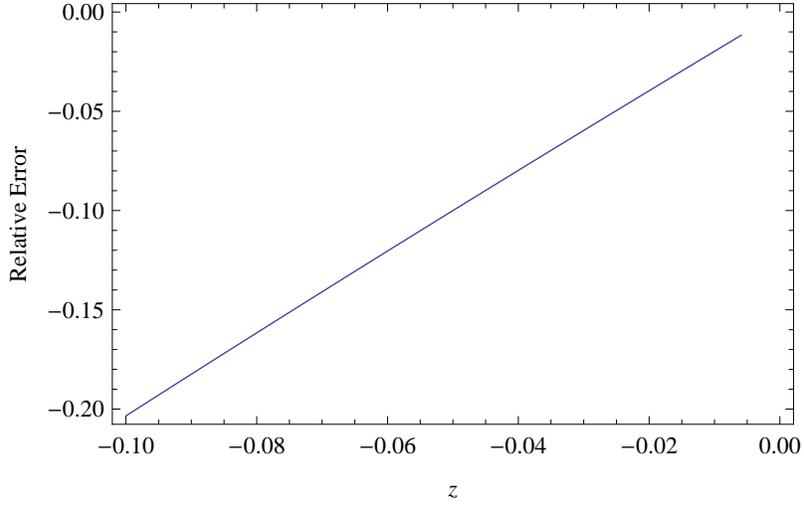}
\caption{\label{fig2}
Relative error of the asymptotic approximation (\ref{asym}).}
\end{figure}

To find the energy density for the region to the left of the wall, $z<0$, 
we  solve Eq.~(\ref{gfe}) in the two regions, always assuming $z'<0$,
in terms of the variable $\kappa^2=k^2+\zeta^2$:
\begin{subequations}
\bea
z<0:\quad g(z,z')&=& \frac1{2\kappa}e^{-\kappa|z-z'|}+A(z') e^{\kappa z},\\
z>0:\quad g(z,z')&=& B(z') \mbox{Ai}(\kappa^2+z).
\eea
\end{subequations}
Here we have chosen the boundary conditions that as $z\to\pm\infty$, 
the Green's function must vanish.
The functions $A$ and $B$ are determined by the requirement that the 
function and
its derivative must be continuous at $z=0$.  This leads to two equations
\begin{subequations}
\bea
B(z')\mbox{Ai}(\kappa^2)&=&\frac1{2\kappa}e^{\kappa z'}+A(z'),\\
\frac1\kappa B(z')\mbox{Ai}'(\kappa^2)&=&-\frac1{2\kappa}e^{\kappa z'}+A(z'),
\eea
\end{subequations}
which may be immediately solved:
\begin{subequations}
 \bea
A(z')&=&\frac1{2\kappa}e^{\kappa z'}\frac{1+\mbox{Ai}'(\kappa^2)/\kappa 
\mbox{Ai}(\kappa^2)}
{1-\mbox{Ai}'(\kappa^2)/\kappa \mbox{Ai}(\kappa^2)},\\
B(z')&=&\frac{e^{\kappa z'}}{\kappa\mbox{Ai}(\kappa^2)-\mbox{Ai}'(\kappa^2)}.
\eea
\end{subequations}
Thus, in particular, the reduced Green's function in the potential-free 
region is
\bea
z,z'<0:\quad g(z,z')&=&\frac1{2\kappa}e^{-\kappa|z-z'|}+\frac1{2\kappa} 
e^{\kappa(z+z')}
\frac{1+\mbox{Ai}'(\kappa^2)/\kappa \mbox{Ai}(\kappa^2)}
{1-\mbox{Ai}'(\kappa^2)/\kappa \mbox{Ai}(\kappa^2)}.
\eea
When we insert this into the expression for the energy density (\ref{u}) we 
omit the
vacuum term in the Green's function, since that has no knowledge of the 
potential, and was completely analyzed in the previous section.  We
are left with for $z<0$ ($|z|\gg\tau$)
\be
u(z)-u_0=\frac{1-6\xi}{6\pi^2}\int_0^\infty d\kappa\,\kappa^3 e^{2\kappa z} 
\frac{1+\mbox{Ai}'(\kappa^2)/\kappa \mbox{Ai}(\kappa^2)}
{1-\mbox{Ai}'(\kappa^2)/\kappa \mbox{Ai}(\kappa^2)}.\label{uleft}
\ee
Unlike the integral over real phase shifts \cite{fulling}, 
the integrand is monotonically tending to zero as $\kappa\to\infty$.
The integral is therefore finite for all $z<0$, 
and may be very easily evaluated
by Mathematica. The results are shown in Fig.~\ref{fig1}.  It is seen that the
energy density diverges as $z\to0$, not at $z=1$; in fact, by 
using the asymptotic expansion of the Airy function,
\be
\frac{1+\mbox{Ai}'(\kappa^2)/\kappa \mbox{Ai}(\kappa^2)}
{1-\mbox{Ai}'(\kappa^2)/\kappa \mbox{Ai}(\kappa^2)}\sim -\frac1{8\kappa^3},
\quad\kappa\to\infty,\label{ratiol}
\ee
 it behaves for small negative $z$ like
\be
u\sim \frac{1-6\xi}{96\pi^2}\frac1z.\label{asym}
\ee
The comparison with the exact numerical integration with this leading 
asymptotic behavior is also shown in Figs.~\ref{fig1}, \ref{fig2}.


The solution for the Green's function inside the wall is
\bea
0<z,z':\quad g(z,z')=\pi\mbox{Ai}(\kappa^2 +z_>)\mbox{Bi}(\kappa^2+z_<)
-\frac{(\kappa\mbox{Bi}-\mbox{Bi}')(\kappa^2)}
{(\kappa\mbox{Ai}-\mbox{Ai}')(\kappa^2)}\pi \mbox{Ai}(\kappa^2+z)
\mbox{Ai}(\kappa^2+z').\nn\\\label{airygf}
\eea
The energy density within the wall is given by Eq.~(\ref{ed}), or
\bea
u&=&\frac1{8\pi^2}\int_0^\infty d\kappa\,\kappa^2\int_{-1}^1 d\cos\theta\bigg\{
\left[\kappa^2+2\frac{\partial^2}{\partial \tau^2}+z\right] g(z,z)\nn\\
&&\quad\mbox{}+\frac{\partial}{\partial z}\frac{\partial}{\partial z'} g(z,z')\big|_{z'\to z}
-2\xi\frac{\partial^2}{\partial z^2} g(z,z)\bigg\}e^{i\kappa\tau\cos\theta}.
\label{genu}\eea
Because both terms in $g$ involve Airy functions of argument $\kappa^2+z$,
we can use the differential equation for the Airy function to write the above
as
\be
u=\frac1{8\pi^2}\left[(1-4\xi)\frac{\partial^2}{\partial z^2}
+4\frac{\partial^2}
{\partial \tau^2}\right]\int_0^\infty d\kappa\,\kappa \,g(z,z)
\frac{\sin\kappa\tau}\tau.\label{genutoo}
\ee

Let us analyze the divergence structure, by considering 
the first term in $g$, Eq.~(\ref{airygf}), which would be the term arising if the
linear potential existed over all space, because $\mbox{Ai}(z)\to 0$ as
$z\to\infty$, while $\mbox{Bi}(z)\to 0$ as $z\to-\infty$.
In any case, this term corresponds to  the
bulk energy density
\be
\tilde u_0=
\frac1{8\pi^2}\left[(1-4\xi)\frac{\partial^2}{\partial z^2}+4\frac{\partial^2}
{\partial \tau^2}\right]\int_0^\infty d\kappa\,\kappa
\pi\mbox{Ai}(\kappa^2+z)\mbox{Bi}(\kappa^2+z)\frac{\sin\kappa\tau}\tau.
\label{bulk}
\ee
To see the divergence structure, use the leading asymptotic behavior
\be
\pi \mbox{Ai}(\kappa^2+z)\mbox{Bi}(\kappa^2+z)\sim \frac12\frac1{
\sqrt{\kappa^2+z}},\label{asylin}
\ee
for large $\kappa$. Then we write the resulting $\kappa$ integral as
\be
\int_0^\infty d\kappa\,\kappa \frac1{\sqrt{\kappa^2+z}}\sin\kappa\tau
=\sqrt{z}\int_0^\infty dy\sin\left(\sqrt{y^2-1}\sqrt{z}\tau\right),
\ee
which for small $\tau$ is dominated by large $y$, so that the integral
can be approximated by
\be
\sqrt{z}\int_1^\infty dy\left\{\left(1-\frac{z\tau^2}{8y^2}\right)
\sin y\sqrt{z}\tau
-\frac{\sqrt{z}\tau}{2y}\left(1+\frac1{4y^2}\right)\cos y\sqrt{z}\tau\right\}.
\label{linint}
\ee
The required integrals are, for small $\beta$,
\begin{subequations}
 \bea
\int_1^\infty dy\sin\beta y&=&\frac1\beta,\quad
\int_1^\infty dy\frac{\sin\beta y}{y^2}=-\beta\ln\beta +\order(\beta),
\label{betaint1}\\
\int_1^\infty dy\frac{\cos\beta y}y&=&-\ln\beta+\mbox{constant},
\quad \int_1^\infty dy\frac{\cos\beta y}{y^3}=\frac{\beta^2}2\ln\beta
+\order(\beta^2)
\eea
\end{subequations}
and then we see only the $\tau$ derivative term contributes in 
Eq.~(\ref{bulk}), and
we obtain the expected result \cite{fulling}
\be
\tilde u_0\sim \frac3{2\pi^2}\frac1{\tau^4}-\frac{z}{8\pi^2\tau^2}
+\frac{z^2}{32\pi^2}\ln\tau,\label{w1}
\ee
as the cutoff $\tau\to0$.  (Our point-splitting procedure would probably not
reveal a possible $\delta$-function
contribution suggested in Ref.~\cite{fulling}.)

\section{General $z^\alpha$ potential}
In general, for an $\alpha$ wall, described by the potential
\be
v(z)=\left\{\begin{array}{cc}
0,&z<0,\\
z^\alpha,& z>0,
\end{array}\right.
\ee
with $\alpha>0$, we construct the reduced Green's function in terms of
the two independent solutions in the region of the potential
\be
\left(-\frac{\partial^2}{\partial z^2}+\kappa^2+z^\alpha\right)\left\{
\begin{array}{c}F(z)\\G(z)\end{array}\right.=0,\label{eqfg}
\ee
where $F(z)$ is chosen to vanish as $z\to+\infty$, and $G(z)$ is an
arbitrary independent solution.  The Wronskian is
\be
w=F(z)G'(z)-G(z)F'(z),
\ee
which is just a constant.

The Green's function to the left of the wall is
\be
z,z'<0,\quad g(z,z')=\frac1{2\kappa}e^{-\kappa|z-z'|}+\frac1{2\kappa}
e^{\kappa(z+z')}\frac{F(0)+F'(0)/\kappa}{F(0)-F'(0)/\kappa},
\ee
and to the right of the wall,
\be
z,z'>0:\quad g(z,z')=\frac1 w F(z_>)G(z_<)-\frac1w F(z)F(z')\frac{G(0)
-G'(0)/\kappa}{F(0)-F'(0)/\kappa}.
\ee  (Adding an arbitrary multiple of $F$ to $G$, of course, leaves this
expression unchanged.)

For $\alpha=1$, $F(z)=\mbox{Ai}(\kappa^2+z)$, $G(z)=\mbox{Bi}(\kappa^2+z)$,
and $w=1/\pi$, and we recover the result in the previous section.
For $\alpha=2$, $F(z)=U(\kappa^2/2,\sqrt{2}z)$, 
$G(z)=U(\kappa^2/2,-\sqrt{2}z)$, in terms of the parabolic cylinder function
\cite{NIST,Bender}.  Alternative notations for this function are
\be
U(a,x)=D_{-a-1/2}(x).
\ee
The value of the parabolic cylinder function, and its derivative,
 at the origin is
\begin{subequations}
\bea
D_\nu(0)&=&\sqrt{\pi}2^{\nu/2}/\Gamma(1/2-\nu/2),\\
D'_\nu(0)&=&-\sqrt{\pi}e^{\nu/2+1/2}/\Gamma(-\nu/2).
\eea
\end{subequations}
Therefore, the Wronskian is
\be
w=\frac{\pi 2^{3/2-\kappa^2/2}}{\Gamma(\kappa^2/4+1/4)\Gamma(\kappa^2/4
+3/4)}.
\ee

\begin{figure}
\centering
\includegraphics{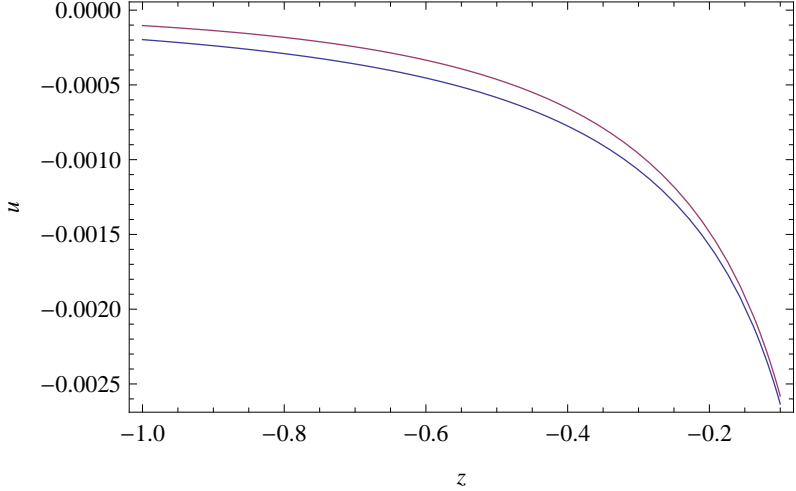}
\caption{\label{quad}
The lower curve shows the exact energy density for the quadratic wall,
for $z<0$, using Eq.~(\ref{ulquad}) with Eq.~(\ref{coefquad}).  The upper
curve is the asymptotic approximation to that energy density, given
by Eq.~(\ref{uquada}).  Again the factor $6\xi-1$ is divided out.}
\end{figure}

The energy density to the left of the wall, $z<0$, is immediately
generalized from Eq.~(\ref{uleft}):
\be
u(z)-u_0=\frac{1-6\xi}{6\pi^2}\int_0^\infty d\kappa\,\kappa^3\,e^{2\kappa z}
\frac{F(0)+F'(0)/\kappa}{F(0)-F'(0)/\kappa}.\label{ulquad}
\ee
For the quadratic wall
\be
\frac{F(0)+F'(0)/\kappa}{F(0)-F'(0)/\kappa}=\frac{1-\frac2\kappa
\frac{\Gamma(\kappa^2/4+3/4)}{\Gamma(\kappa^2/4+1/4)}}
{1+\frac2\kappa
\frac{\Gamma(\kappa^2/4+3/4)}{\Gamma(\kappa^2/4+1/4)}}.\label{coefquad}
\ee
Asymptotically,
\be
\frac{\Gamma(\kappa^2/4+3/4)}{\Gamma(\kappa^2/4+1/4)}\sim\frac\kappa2\left(
1+\frac1{4\kappa^4}\right),\quad\kappa\to\infty,\label{ratioq}
\ee
so we approximate the exact energy density to the left of the wall by
\be
u(z)-u_0\sim -\frac{1-6\xi}{6\pi^2}\int_1^\infty \frac{d\kappa}{8\kappa} 
e^{2\kappa z}=-\frac{1-6\xi}{48\pi^2}\Gamma(0,-2z),\label{uquada}
\ee
in terms of the incomplete gamma function.  The latter is actually a very
accurate approximation as Fig.~\ref{quad} shows.

In the region of the potential, we can calculate the generalization of the
``bulk energy'' (\ref{bulk}),
\be
\tilde u_0=\frac1{8\pi^2}\left[(1-4\xi)\frac{\partial^2}{\partial z^2}
+4\frac{\partial^2}{\partial\tau^2}\right]\int_0^\infty d\kappa\,\kappa\,
\tilde g_0(z,z)\frac{\sin\kappa\tau}{\tau},\label{bulk2}
\ee
because the argument leading from Eq.~(\ref{genu}) to Eq.~(\ref{genutoo}) 
holds for an arbitrary potential. Here, for the quadratic wall,
\be
\tilde g_0(z,z')=\frac1w U(\kappa^2/2,\sqrt{2}z)U(\kappa^2/2,-\sqrt{2}z).
\ee
The uniform asymptotic approximation for large order for $U(\kappa^2/2,
\sqrt{2}\kappa t)$ is given in the NIST handbook \cite{NIST}.  The leading
approximation is rather immediately found to yield
\be
\tilde g_0(z,z)\sim\frac1{2\sqrt{\kappa^2+z^2}}
-\frac18\frac1{(\kappa^2+z^2)^{5/2}}+\dots,\label{laquad}
\ee
where the subleading term is explained in the following.
The leading term differs from Eq.~(\ref{asylin}) simply
 by changing the potential from $z$ to $z^2$.
This means that we can make the same substitution in the integral
(\ref{linint}), and so the bulk energy density (\ref{bulk2}) is 
\be
\tilde u_0=\frac3{2\pi^2}\frac1{\tau^4}-\frac1{8\pi^2}\frac{z^2}{\tau^2}
+\frac1{32\pi^2}\left[z^4+2(1-4\xi)-\frac23\right]\ln\tau,\label{w2}
\ee
where the $-(2/3)\ln\tau$ term arising from the subleading term in
Eq.~(\ref{laquad}) results in the appearance of the conformal
coefficient $(4/3)(1-6\xi)$ for the constant term multiplying $\ln\tau$.
This last result may be easily generalized to an arbitrary
potential $v(z)$.  The bulk Green's function at
coincident points can be written
as
\be
\tilde g_0(z,z)=\frac1{G'(z)/G(z)-F'(z)/F(z)}.
\ee
The leading asymptotic behavior of the solutions is
given by the WKB approximation \cite{Bender},
\begin{subequations}
\bea
F(z)&\sim& Q^{-1/4}(z)\exp\left[-\int^z dt\left(Q^{1/2}(t)
+\frac{v''(t)}{8Q^{3/2}(t)}\right)
\right],\\
G(z)&\sim& Q^{-1/4}(z)\exp\left[\int^z dt\,\left(Q^{1/2}(t)
+\frac{v''(t)}{8Q^{3/2}(t)}\right)\right],
\eea
\end{subequations}
where $Q(z)=\kappa^2+v(z)$.  Here it was necessary to keep the first
subleading correction, as given in Ref.~\cite{Bender}.
Thus, for large $\kappa$,
\be
\frac{G'(z)}{G(z)}-\frac{F'(z)}{F(z)}\sim 2 Q^{1/2}(z)\left(1+\frac{v''(z)}
{8Q^2(z)}\right).
\ee
This is the immediate generalization of Eqs.~(\ref{asylin}) and (\ref{laquad}).
Then, the generalization of the Weyl expansion (\ref{w1}) and (\ref{w2}) 
is\footnote{Ref.~\cite{fulling} proposes that these potential divergences
could be subtracted by renormalizing terms in the Lagrangian describing
the background field $v$.} 
\be
\tilde u_0\sim \frac{3}{2\pi^2}\frac1{\tau^4}-\frac1{8\pi^2}\frac{v}{\tau^2}
+\frac1{32\pi^2}\left[v^2+\frac23
(1-6\xi)\frac{\partial^2}{\partial z^2}v\right]
\ln\tau,
\ee
which uses the evaluation
\be
\int_0^\infty \frac{d\kappa\,\kappa}{(\kappa^2+v)^{5/2}}\sin\kappa\tau
\sim\frac16\tau^3\ln\tau,
\ee
which follows from Eq.~(\ref{betaint1}).  Note that the derivative term
vanishes for the conformal value of $\xi$.  This form, of course, follows
from the general heat kernel consideration of this problem, and is seen
for $\xi=1/4$ in Ref.~\cite{fulling}.

The behavior of the energy density to the left of the wall is also worked out
easily, in general.  We rescale the equation (\ref{eqfg}) for $F(z)$, so that
$z=x/\kappa$,
\be
\left(-\frac{d^2}{dx^2}+1+\kappa^{-2-\alpha}x^\alpha\right)F(z)=0,
\ee
and then solve this equation perturbatively in powers of $\kappa^{-1}$.
$F$ has the form
\be
F(x) \sim e^{-x}\left(1+\frac1{\kappa^\beta}f(x)+\dots\right),
\quad\kappa\to\infty,\label{F}
\ee
where consistency requires $\beta=2+\alpha$, and $f$ satisfies
\be
f''(x)-2f'(x)-x^\alpha=0.
\ee
This may be immediately solved for $f'$,
\be
f'(x)=e^{2x}\left(\int_0^x dt\,t^\alpha \,e^{-2t}+C\right),
\ee
in terms of a constant $C$.  This will reverse the required decreasing
exponential dependence seen in Eq.~(\ref{F}) unless
\be
C=-\int_0^\infty dt\,t^\alpha\,e^{-2t}=-2^{-1-\alpha}\Gamma(1+\alpha).
\ee
In particular, this determines
\be
f'(0)=-\frac{\Gamma(1+\alpha)}{2^{1+\alpha}}.
\ee
From this we determine the required ratio occurring in Eq.~(\ref{ulquad})
\be
\frac{F'(z=0)}{\kappa F(z=0)}=\frac{F'(x=0)}{F(x=0)}\sim-1
+\frac1{\kappa^{2+\alpha}}f'(0)
=-1-\frac{\Gamma(1+\alpha)}{2^{1+\alpha}\kappa^{2+\alpha}},
\ee
or
\be
\frac{1+F'(0)/\kappa F(0)}{1-F'(0)/\kappa F(0)}\sim
-\frac{\Gamma(1+\alpha)}{(2\kappa)^{2+\alpha}},
\ee
which generalizes Eqs.~(\ref{ratiol}) and (\ref{ratioq}).
This gives the asymptotic estimate for the energy density near the wall on
the left:
\be
u(z)-u_0\sim -\frac{1-6\xi}{96\pi^2}|z|^{\alpha-2}\Gamma(1+\alpha)
\Gamma(2-\alpha,2|z|),\quad z\to0-,\label{genzto0}
\ee
which generalizes Eqs.~(\ref{asym}) and (\ref{uquada}).  The singularity at 
$z=0$ disappears for $\alpha>2$; 
\be
u(0)-u_0=\frac{1-6\xi}{96\pi^2}\frac{\Gamma(1+\alpha)}{2-\alpha}2^{2-\alpha},
\quad\alpha>2.
\ee
For $\alpha<2$,
\be u(z)-u_0=-\frac{1-6\xi}{96\pi^2}\Gamma(1+\alpha)\left(|z|^{\alpha-2}
\Gamma(2-\alpha)-\frac{2^{2-\alpha}}{2-\alpha}\right),
\ee
which as $\alpha\to 2$ from below approaches
\be
u(z)-u_0=\frac{1-6\xi}{48\pi^2}\left(\gamma+\ln2|z|\right),
\ee
an accurate approximation to the general estimate (\ref{genzto0}).
\section{Conclusions}

We have explored in this paper the nature of the divergences that occur in 
the energy density in quantum field theory near walls, for the case of 
scalar fields. We generalize the walls from being perfect Dirichlet 
boundaries, to potentials of the form $z^\alpha$ within the region of
the infinite wall. Besides the usual Weyl volume divergence, which arises 
from the free part of the theory, the energy density exhibits a divergence 
as the wall is approached if the wall is not too soft, $\alpha\le2$.  
That divergent term, however, vanishes if the conformal stress tensor, 
characterized by $\xi=1/6$, is used.  Correspondingly, there is no observable
consequence of this surface-divergent term, absent gravity. We also compute 
the divergences that occur within the region of the wall, which depend
on the form of the potential.  To obtain unambiguously observable consequences
we would need to consider the interaction between two such walls.

A question arises as to how seriously to take the cutoff.  As we noted
for the Dirichlet wall, if the form for the energy density is taken
literally for $z<\tau$, we obtain a nonvanishing, $\xi$-independent result 
for the energy of a single wall, agreeing with the expected area term
in the Weyl expansion, Eq.~(\ref{W2}).

How is this analysis generalized for more realistic theories?  A similar 
divergence in the energy density occurs near a perfectly conducting boundary 
if one considers only the electric or the magnetic part of the energy, or 
the TE and TM modes separately \cite{milton10}.  The results there were
for parallel plates separated by a distance $a$, so if we take $a\to\infty$
there we recover the energy density for a single conducting wall.
The electric and magnetic  energy densities near such a perfect boundary
at $z=0$ are (the volume divergence is omitted here)
\be u_E(z)=-u_M(z)=
\int\frac{d\zeta (d\mathbf{k_\perp})}{(2\pi)^3}\frac\kappa2 e^{-2\kappa z}
e^{i\zeta\tau},\quad z>0,
\ee
again keeping the point-splitting regulator.  Carrying out the integration,
we find
\be
u_E(z)=-u_M(z)=
\int_0^\infty \frac{d\kappa\,\kappa^3}{8\pi^2}e^{-2\kappa z}\frac
{2\sin\kappa\tau}{\kappa\tau}=\frac1{2\pi^2}\frac{\tau^2-12 z^2}
{(\tau^2+4z^2)^3}.\label{ueum}
\ee  If the regulator is removed, $\tau\to0$,
\be
u_E=-u_M\to -\frac3{32\pi^2 z^4},
\ee
the same type of quartic divergence encountered in the nonconformal scalar
case, Eq.~(\ref{dsdiv}).   This result was first observed by Dewitt 
\cite{dewitt} more than 35 years ago.
 Not only does this energy cancel when the electric and magnetic terms
are combined, but if this energy density is integrated over all space to
the right of the plate,
\be
\int_0^\infty dz \,u_E(z)=\frac1{2\pi^2}\int_0^\infty dz
\frac{\tau^2-12 z^2}{(\tau^2+4z^2)^3}=0,
\ee
we get a vanishing energy! [This result also follows from integrating
the integral form of Eq.~(\ref{ueum}) over $z$ and using Eq.~(\ref{fints}).]
 So these surface divergences have but an ephemeral existence.  
(These cancellations do not occur, however, for dielectric
interfaces \cite{milton10}.)

If we wish to examine the surface divergences in the complete
stress-energy tensor in the electromagnetic case, it is better, of
course, to break up the description into TE and TM modes.  For such
a local description, we need the rotationally invariant form of the
electromagnetic Green's dyadic given in Ref.~\cite{qfext}.  Then, it
is a straightforward calculation using the methods described in this
paper to obtain the stress tensor for the TE and TM modes in the presence
of a perfectly conducting wall at $z=0$, for $|z|\gg \tau$:
\be
\langle t^{\mu\nu}_{\rm{TE,TM}}\rangle =\frac1{2\pi^2\tau^4}\mbox{diag}(3,1,1,1)
\mp\frac1{32\pi^2 z^4}\mbox{diag}(2,1,1,0).
\ee
Remarkably, both terms have vanishing trace, so the individual modes
respect conformal symmetry even in the presence of the wall.  The
$z$-dependent surface term cancels for the complete electromagnetic
contribution. Neither term would seem to have any observable consequence.

\acknowledgments
This work was supported in part by grants from the National Science Foundation 
and the US Department of Energy.  
I am grateful for numerous conversations with Steve Fulling, and to Klaus
Kirsten for useful comments.  I thank Elom Abalo, Prachi Parashar, Nima Pourtolami, 
and Jef Wagner for collaborative assistance.

\end{document}